\documentclass[twocolumn,aps,prb,showpacs,superscriptaddress,amsfonts,amssymb,amsmath]{revtex4}

\begin{document}

\draft
\title{Anomalous Kondo effect of indirectly coupled double quantum dots}

\author{Qing-feng Sun$^{1,2}$, Hong Guo$^{2}$, and Yupeng Wang$^1$}

\address{
$^1$Institute of Physics, Chinese Academy of Sciences,
Beijing 100080, China\\
$^2$Center for the Physics of Materials and Department
of Physics, McGill University, Montreal, PQ, Canada H3A 2T8}

\begin{abstract}

We report theoretical investigations of indirectly coupled double quantum
dots (QD) side connected to an one-dimensional quantum wire. Due to quantum
interference controlled by the parameter $k_F L$, with $k_F$ the Fermi
wave number of the wire and $L$ the distance between the two QDs, distinctly
different Kondo resonances are predicted depending on the range of $k_F L$.
A true bound Kondo states is found while an anomalous Kondo resonance gives
rise to both reduction and enhancement of conductance.

\end{abstract}

\maketitle

In the original Kondo physics concerning magnetic impurities in
metal\cite{kondo1}, effects of multiple impurities have been subjected to
extensive study dating back to the 1960's\cite{ref7}. Impurity
scattering is characterized by a parameter $k_F l_m$ where $k_F$ is the
Fermi wavevector of the host metal and $l_m$ is the mean momentum relaxation
length\cite{ref7}. Impurities in a metal are situated randomly and typically
one studies an {\it average} physical behavior. The recent extensive
investigations of Kondo phenomenon in quantum dot (QD) devices, both
theoretically and experimentally\cite{ref1,ref2}, is due partly to the
fact that such an important and prototypical many-body physics can be
systematically investigated using QDs whose parameters are precisely
controllable. However, despite the numerous investigations of QD Kondo effect,
to the best of our knowledge, the QD parameter playing a similar role as the
$k_F l_m$ of a metal, has not yet been explored.

In this paper we report a theoretical investigation of Kondo effect in a
device where two QDs side-couple to the same quantum wire but otherwise no
direct coupling of any sort between them, as shown in Fig.(1a). Of particular
interest is the dimensionless device parameter $k_F L$ where $k_F$ is the
Fermi wavevector of the quantum wire and $L$ the distance between the two
QD's.
Clearly, $k_F L$ plays the role of $k_F l_m$ of a metal, except that it can
be precisely controlled to give unambiguous results about strongly correlated
effects in quantum transport at reduced dimensions. There are several
theoretical studies of Kondo physics in two QDs connected in
series\cite{ref8,ref9}, but systems studied so far all involve direct QD-QD
coupling either through a tunnel coupling between the QD's or/and through the
QD-QD Coulomb interaction. Experimentally, double-QD Kondo physics also
starts to attract attention\cite{ref10}. The present work will focus on the
role of $k_F L$.

For a single QD weakly coupled to two leads in the form of Lead-QD-Lead,
the signature of Kondo effect is the {\it enhancement} of device
conductance\cite{ref4} as temperature ${\cal T}$ is lowered to below the
Kondo temperature $T_K$. On the other hand, for a single QD side-coupled
to a quantum wire, the signature of Kondo effect is the {\it reduction}
of conductance of the quantum wire\cite{ref5}. The QD Kondo effect is
due to quantum superposition of electron co-tunneling events\cite{kondo1}
which is illustrated by the energy diagram shown in Fig.(1b) for a
side-coupled single QD, where two electrons with different spins
tunnel in/out of the QD in very short time scales. Below $T_K$, the Kondo
effect makes the QD an efficient scatterer for electrons traversing the
quantum wire thereby reducing conductance while screening spin of the QD.

For the device we consider (Fig.(1a)), although there is no direct
QD-QD coupling, an electron may still flow from a QD to the other
through the quantum wire, {\it i.e.} there is an {\it indirect}
coupling between the QDs controlled by the parameter $k_F L$. Due
to multiple reflections between the two QDs, quantum interference
effect is found to give rise to a qualitatively different Kondo
behavior in this device as compared to that of a single
side-coupled QD system\cite{ref5}. In particular, we found that a
most general behavior for this system is that physical quantities
such as local density of states (LDOS), transmission coefficient
$T(E)$, and conductance $G$, are all periodic functions of
parameter $k_F L$ with a period of $\pi$. The Kondo effect is now
separated into two distinct regimes given by $k_F L < \pi/2$ and
$k_F L > \pi/2$. For $0<k_F L <\frac{\pi}{2}$, Kondo resonance can
both reduce or enhance conductance of the quantum
wire---qualitatively different from a single QD
device.\cite{ref4,ref5} On the other hand for $\pi/2 < k_F L <
\pi$, Kondo effect can only reduce conductance. Exactly at $k_F
L=\pi$, a true bound Kondo state is established in the region of
the two QDs such that transport along the wire is enhanced
at much lower temperatures.

The system of Fig.(1a) can be described by the Anderson Hamiltonian (not
shown). For simplicity we consider a single level $\epsilon_j$ ($j=1,2$
labels the QD) in each QD with spin index $\sigma$ ($N=2$ fold degenerate),
and a constant on-site Coulomb interaction $U$. In the
$U\rightarrow \infty$ limit, double occupancy of $\epsilon_j$ becomes
unlikely therefore the on-site interaction does not occur, hence one
can neglect $U$ but enforcing the single occupancy of $\epsilon_j$.
This idea is realized in the slave-boson approach\cite{ref8,ref11} which we
will use, and our results are qualitatively reasonable in the large $U$ limit.
In the slave-boson representation\cite{ref8,ref11}, the Hamiltonian of the
Anderson model describing the device is then transformed
to the following form:\cite{ref8,ref12}
\begin{eqnarray}
H & = & \sum\limits_{k\sigma} \epsilon_k c_{k\sigma}^{\dagger} c_{k\sigma}
  +\sum\limits_{\sigma j} \frac{t_j}{\sqrt{N}}
   \left[ c^{\dagger}_\sigma(a_j) b^{\dagger}_j f_{j\sigma} +H.c. \right]
   \nonumber \\
& +& \sum\limits_{\sigma j} \epsilon_j f^{\dagger}_{j\sigma} f_{j\sigma}
  +\sum\limits_j \lambda_j \left[ b_j^{\dagger} b_j
   +\sum\limits_{\sigma} f^{\dagger}_{j\sigma} f_{j\sigma} -1
   \right]
\label{hamiltonian}
\end{eqnarray}
where $c^{\dagger}_{k\sigma}$($c_{k\sigma}$) is the electron
creation (annihilation) operator in the one-dimension quantum wire
with corresponding energy $\epsilon_k =\frac{\hbar^2 k^2}{2 m}
+U_0$, here $m$ is the effective mass of the electron and $U_0$
the potential energy in the quantum wire. $b_j$ and $f_{j\sigma}$
are the slave-boson operator and the pseudofermion operator,
respectively. The last term represents the single-occupancy
constraint $b_j^{\dagger} b_j + \sum\limits_{\sigma}
f^{\dagger}_{j\sigma} f_{j\sigma} =1$ in each QD with Lagrange
multiplier $\lambda_j$. The two QDs are side-coupled to the
quantum wire at positions $x=a_j$ ($j=1,2$), with $t_j$ the
tunneling matrix element. Finally, the annihilation operator
$c_{\sigma}(x)$ at position $x$ is given by $c_{\sigma}(x) = \int
\frac{dk}{\sqrt{2\pi}} c_{k\sigma} e^{ikx} $. In the following, we
adopt a mean field theory in which the slave-boson operator ${\hat
b}_j$ is taken as a constant c-number which is solved
self-consistently. The mean field theory is a qualitative correct
for describing Kondo regime at low temperature with zero bias
voltage.\cite{ref8,ref12}

To obtain conductance $G$ and intradot LDOS, we calculate the
retarded Green's functions $G^r$ for Hamiltonian
(\ref{hamiltonian}) from their definitions, $G^r(xt, x'0)\equiv
-i\theta(t)<\{ c_{\sigma}(xt),c_{\sigma}^{\dagger}(x0)\}>$ and
$G^r_{ij}(t,0) \equiv -i \theta(t) <\{ f_i(t), f_j^{\dagger}(0)
\}>$, where continuous variables $x,x'$ are for the quantum wire
and discrete indices $i,j$ for the two QDs. If the QDs and the
quantum wire do not couple ($t_1=t_2=0$), these Green's functions
can be solved exactly and we denote them by $g^r(x,x',E)$ and
$g^r_{ij} (E)$:\cite{ref13} $g^r(x,x',E)=-i\pi \rho(E)
e^{ik|x-x'|} $, $g^r_{ij} (E) =\delta_{ij} /(E-\tilde{\epsilon}_j
+i 0^+) $, where all time dependence have been Fourier transformed
into energy $E$ dependence. Here $k=\sqrt{2m(E-U_0)}/\hbar$,
$\tilde{\epsilon}_j =\epsilon_j +\lambda_j$, and LDOS of the wire
is $\rho(E) =\frac{m}{\pi \hbar^2 k}$. Using $g^r(x,x',E),g^r_{ij}
(E)$ as unperturbed Green's function and apply Dyson equation with
terms involving $t_j$ as the potential, we obtain exactly
\begin{eqnarray}
& &
G^r(x,x',E)= g^r(x,x')+\sum\limits_{i,j} g^r(x,a_i)\tilde{t}_i
            G^r_{ij} \tilde{t}_j g^r(a_j,x')  \nonumber \\
& &
G^r_{jj}(E) =\frac{ g^{r-1}_{\bar{j}\bar{j}} -\Sigma^r_{\bar{j}}}
     {(g^{r-1}_{j j} -\Sigma^r_{j})
      (g^{r-1}_{\bar{j}\bar{j}} -\Sigma^r_{\bar{j}})
      -\Sigma^r_j \Sigma^r_{\bar{j}} e^{2ikL} } \\
& & G^r_{j\bar{j}} (E)=
 \tilde{t}_j g^r(a_j,a_{\bar{j}})\tilde{t}_{\bar{j}} G^r_{\bar{j}\bar{j}}/
 (g^{r-1}_{j j} -\Sigma^r_j) \nonumber
\end{eqnarray}
where ${\bar{j}}=2$ if $j=1$, or ${\bar{j}}=1$ if $j=2$;
$\tilde{t}_j =\frac{t_j b_j}{\sqrt{N}}$; and $L=|a_1-a_2|$. The self-energy
$\Sigma^r_j(E) = \tilde{t}^2_j g^r(a_j,a_j,E) \equiv -\frac{i}{N}
\Gamma_j(E) b_j^2 \equiv -i \tilde{\Gamma}_j$. From $G^r$, the transmission
coefficient $T(E)$, conductance $G$, and intradot $LDOS_j(E)$ can all be
calculated\cite{ref13} straightforwardly:
$T(E)=|i\hbar v G^r(a_1,a_2,E)|^2$,
$G=\frac{2e^2}{h} \int dE T(E)
\left(\frac{-\partial f(E)}{\partial E}\right) $,
and $LDOS_j(E) = -\frac{1}{\pi} Im G^r_{jj}(E)$. Here
$v=\frac{\hbar k}{m}= 1/ \pi \hbar \rho(E)$
is the electron velocity and $f(E)$
the Fermi distribution of the quantum wire. Finally, the above analysis is
supplemented with a self-consistent numerical calculation of the four
unknowns ${b}_j$ and $\lambda_j$. From the constraints and the equation of
motion of the slave-boson operators,\cite{ref8} we derive the self-consistent
equation as: $b^2_j +\sum\limits_{\sigma} n_{j\sigma} =1$ with the intradot
electron occupation number $n_{j\sigma} =\int dE f(E) LDOS_j(E)$ and
$\lambda_j b^2_j = \sum\limits_{\sigma} \int \frac{dE}{\pi}
f(E) Im \left[ g^r_{jj}(E) \tilde{t}_j^2 G^r(a_j,a_j,E)\right] $.

In the following, we fix the chemical potential $\mu$ of the quantum wire
as the energy zero, and assuming that $U_0$ has a large negative value,
$-U_0 \gg max(k_B {\cal{T}}, \Gamma_j, |\epsilon_j|)$, so that
the wave vector $k$ and the density of state $\rho(E)$ can be
approximated as independent of $E$ at $E\sim \mu$. This way
$\Gamma_j(E)$ is also approximately a constant and $k\approx k_F$.
We consider square symmetric bands in which
$\Gamma_j(E)=\Gamma_j \theta(W-|E|)$ and
$W=100 \gg max(k_B {\cal{T}},\Gamma_j, |\epsilon_j|)$. Finally,
we set $\Gamma_1=\Gamma_2\equiv \Gamma=1$ as the energy unit.
We now discuss results.

{\bf LDOS and bound Kondo states.}  Due to multiple reflections
between the two QDs, LDOS, $T(E)$ and $G$ are all periodic
functions of parameter $k_F L$ with a period $\pi$. If each QD
acts independently, then each has its own Kondo resonance at
energy $\tilde{\epsilon}_0$ with a width $\tilde{\Gamma}_0
=\frac{\Gamma b^2}{N}$ and Kondo temperature $k_B T^0_K
=\sqrt{\tilde{\epsilon}_0^2+\tilde{\Gamma}_0^2} =
We^{-\pi|\epsilon_0|/\Gamma}$. With coherence, however, there is a
superposition of these individual Kondo states through multiple
scattering between the two QDs. For example, if the two QDs are
identical ($\epsilon_1=\epsilon_2 \equiv \epsilon$ and
$\tilde{\epsilon}_1 =\tilde{\epsilon}_2 \equiv \tilde{\epsilon}$),
coherent superposition establishes two new Kondo peaks: peak-1, at
energy $(\tilde{\epsilon}+{\tilde{\Gamma}} \sin k_F L)$ with
half-width ${\tilde{\Gamma}} (1 +\cos k_F L)$; and peak-2, at
energy $(\tilde{\epsilon}- {\tilde{\Gamma}} \sin k_F L)$ with
half-width ${\tilde{\Gamma}} (1 -\cos k_F L)$.\cite{note1}. The
thin solid curve (red) in Fig.(1c) plots the LDOS of this
situation at $k_F L=\pi/2$ for which the two Kondo peak heights
are almost equal, while other values of $k_F L$ (dotted and dashed
curves) make unequal peak heights.

A very interesting situation is when $k_F L=n\pi$ where $n$ is an integer:
the width of one of the Kondo resonances vanishes because of the factor
$(1 \pm \cos k_F L)$. This means a true bound Kondo state is localized in
the region of the two QDs within energy continuum. The thick solid curve
(blue) in Fig.(1c) plots the LDOS of the device with $k_F L=0.99\pi$,
showing a very sharp Kondo peak with tiny width. This peak evolves into a
$\delta$-function exactly at $k_F L=\pi$. Fig.(1a) shows the physics behind
the bound Kondo state. During the frequent co-tunneling processes which
gives rise to Kondo effect, an electron can tunnel out of QD-1 and flow
away in the quantum wire, indicated by path-1 (p1) of Fig.(1a).
But this electron can also flow towards QD-2 to participate
co-tunneling there and afterward it flows back, as shown by path-2 (p2).
If $k_F L=n\pi$, path-1 and path-2 will
differ by a phase difference of $(n\pi + \pi + n\pi) = (2n+1)\pi$,
resulting to a destructive interference. It is the quantum
superposition of these destructive interferences which prevents the electron
from flowing away from the two QD region, and therefore a bound Kondo state.
It is very interesting to observe that electrons participating Kondo resonance
can be completely bounded, and this has important effect to the conductance
of the quantum wire (see below).

The above behavior qualitatively holds when the two QDs are not
totally identical. However, as the difference $\Delta \epsilon
\equiv \epsilon_1 -\epsilon_2 $ becomes large, the two QDs act
more independently and the two Kondo peaks gradually localize
themselves into the QDs. With a small but finite $\Delta\epsilon$
and with $k_F L = \pi$, the bound Kondo state develops a tiny
width, much smaller than $k_B T_K^0$, but it now further
coherently superposes with the other Kondo state to produce a
Fano-like interference line shape, as shown in Fig.(1d).

{\bf Transmission coefficient.} Quantum interference also gives
rise to an anomalous behavior in $T(E)$. For a single side-coupled
QD, $T(E)$ should simply decrease when $E$ is near the Kondo
resonance $\tilde{\epsilon}_0$, and reach its minimum $T=0$ for
energy $E=\tilde{\epsilon}_0$ due to Kondo scattering. For our two
QD device, this is still largely true as shown in Fig.(2) where
$T(E)\sim 1$ when $E$ is far away from $\tilde{\epsilon}$ and
$T(E)$ takes small values when $E\sim \tilde{\epsilon}$. However,
due to multiple scattering between the two QDs during the
co-tunneling processes, a resonance transmission become possible
similar to the optical interference of a Fabry-Perot
interferometer. This induces a very sharp peak $T(E)=1$ near
$E\approx 0$ in the otherwise small $T(E)$ background, see
Fig.(2). This resonance peak exists regardless of the value of
$k_F L$, but importantly the peak position $E^*$ crucially depends
on whether or not $k_F L > \pi/2$, as we found
$E^*=\tilde{\epsilon} -\frac{\tilde{\Gamma}}{2}\frac{\sin k_F
L}{\cos k_F L}$. Hence, if $\pi/2 < k_F L < \pi$, $E^*$ is always
positive (i.e. $E^*>\mu$) for any intradot level $\epsilon$ (see
Fig.2d), so that this $T(E^*)$ peak does not contribute to
conductance at Fermi level. On the other hand, for $0 < k_F L <
\pi/2$, $E^*$ can become zero ({\it i.e.} $E^*=\mu$) or negative
which depend on the intradot level $\epsilon$, and the resonance
$T(E^*)$ peak can therefore contribute to Kondo effect. Due to
this $T(E^*)$ peak which originates from quantum interference,
conductance $G$ will also show an anomalous behavior as we show
now.

{\bf Conductance.}  For identical QDs, Fig.(3a,b) plots $G$ versus
the intradot level $\epsilon=\epsilon_1=\epsilon_2$ at low
temperature $\cal{T}$ with different values of $k_F L$. For high
$\epsilon$ ($\epsilon>\mu=0$), $G$ is quite large, $G\sim 2e^2/h$.
This is because transport is not in the Kondo regime and $G$ is
given by the one channel ballistic wire. When $\epsilon$
lowers, $G$ reduces. When $\epsilon < -2\Gamma$ the device is in
the Kondo regime, and $G$ is strongly affected by scattering of
the two side-coupled QDs. For $\frac{\pi}{2}< k_F L < \pi$ (except
$k_F L$ very near $\pi$) , the scattering causes $G$ to strongly
reduce when $\epsilon < -2\Gamma$. Although there is still a peak
in $T(E^*)$ at $E^*>0$ for this range of $k_F L$, this peak plays
no role to Kondo effect as we discussed in the last paragraph.
Hence, qualitatively, the Kondo effect for $\frac{\pi}{2}< k_F L <
\pi$ is similar to that of a single side-coupled QD
device.\cite{ref5} However, the behavior of $G$ becomes
qualitatively different for $0<k_F L <\frac{\pi}{2}$, because the
anomoulous $T(E^*)$ peak will now contribute to Kondo physics. The
striking consequence is that $G$ can now both decrease and
increase in the Kondo regime, the latter is due precisely to the
anomoulous $T(E^*)$ peak. Hence, as shown in Fig.(3a,b), $G$ has
two asymmetric dips and one peak in the region $\epsilon\sim
-2\Gamma$. Letting temperature $\cal{T}$ decrease, the dip becomes
deeper and the peak becomes higher with a slight change of their
positions toward lower $\epsilon$. At last, when intradot level
$\epsilon$ is much lower ({\it e.g.} $\epsilon \sim -4\Gamma$), $G$ can
increase and reach $2e^2/h$ again (see Fig.3b).

Finally, we study the temperature dependence of $G$ (see
Fig.3c,d). At high $\cal{T}$ (${\cal{T}} >T_K^0$), $G\sim 2e^2/h$
for any $k_F L$. Then let $\cal{T}$ reduce, the system gradually
comes into Kondo regime. In the beginning, $G$ decrease for all
$k_F L$ value. However, with further reducing $\cal{T}$, $G$ for
different $k_F L$ shows very different behaviors. For $k_F L
=\frac{\pi}{2}$, $G$ is simply a monotonic decreasing function of
temperature (Fig.3d). On the other hand, for $k_F L \sim \pi$
(i.e. $0$), $G$ becomes non-monotonic as $\cal{T}$, and it
increases again and even can reach $2e^2/h$ at much lower
$\cal{T}$ (see Fig.3c, or the dotted curve in Fig.3a). The
increase of $G$ is surprising since the device is in the Kondo
regime and incident electrons should be back-scattered strongly by
the two side-coupled QDs. This increase of $G$ is due to the
quasi-bound Kondo state for $k_F L \sim \pi$. In other words,
quantum interference due to multiple reflections by the two QDs
enhances the conductance of the wire in the Kondo regime, an
anomalous behavior indeed. It is worth to mention that Fig.3c
clearly shows that $G$ is almost a linearly increasing function of
$log\cal{T}$ at very low $\cal{T}$ (for ${\cal{T}} \sim 10^{-5}$
to $10^{-4}$), indicating a Kondo behavior, {\it i.e.} the very
sharp peak with a tiny width in Fig.1c is indeed from the Kondo
effect, and its Kondo temperature is about $10^{-5}$ for $k_F L
=0.99\pi$ and $\epsilon=-2.1$. If $k_F L$ further approaches to
$\pi$, $T_K$ of the quasi-bound Kondo peak reduces further. On the
other hand, for the other broad Kondo peak (see Fig.1c), its Kondo
temperature is about $10^{-2}$ for the same $k_F L =0.99\pi$ and
$\epsilon=-2.1$, which is three orders of magnitude larger than
the quasi-bound Kondo state.

In summary, we found that the dimensionless parameter $k_F L$ provides a
critical control of the Kondo phenomenon in the double-QD device where the
QDs side-couple to a quantum wire but otherwise no direct coupling between
them. Due to interference of quantum paths of electrons participating
the co-tunneling process, a new Kondo behavior is predicted.
In particular, a true bound Kondo state, an anomalous resonance peak in
transmission coefficient, as well as anomalous conductance behavior, are
found depending on the range of $k_F L$. The predicted phenomena should be
experimentally observable using present technologies. The two side-coupled
QD's device can be fabricated in two-dimensional electron gas (2DEG).
Using Fermi wave length $\lambda_F\sim 50$nm for typical 2DEG, $L\sim 25$nm
gives $k_F L=\pi$ while $L\sim 38$nm gives $k_FL =\frac{3\pi}{2}$, hence
the interesting Kondo effect of this device controlled by parameter $k_F L$
can be accessed experimentally.

{\bf Acknowledgments:}
We gratefully acknowledge financial support from NSERC of Canada, FCAR of
Quebec (Q.S., H.G), the National Science Foundation of China and the Chinese
Academy of Sciences (Q.S. and Y.W).


\newpage

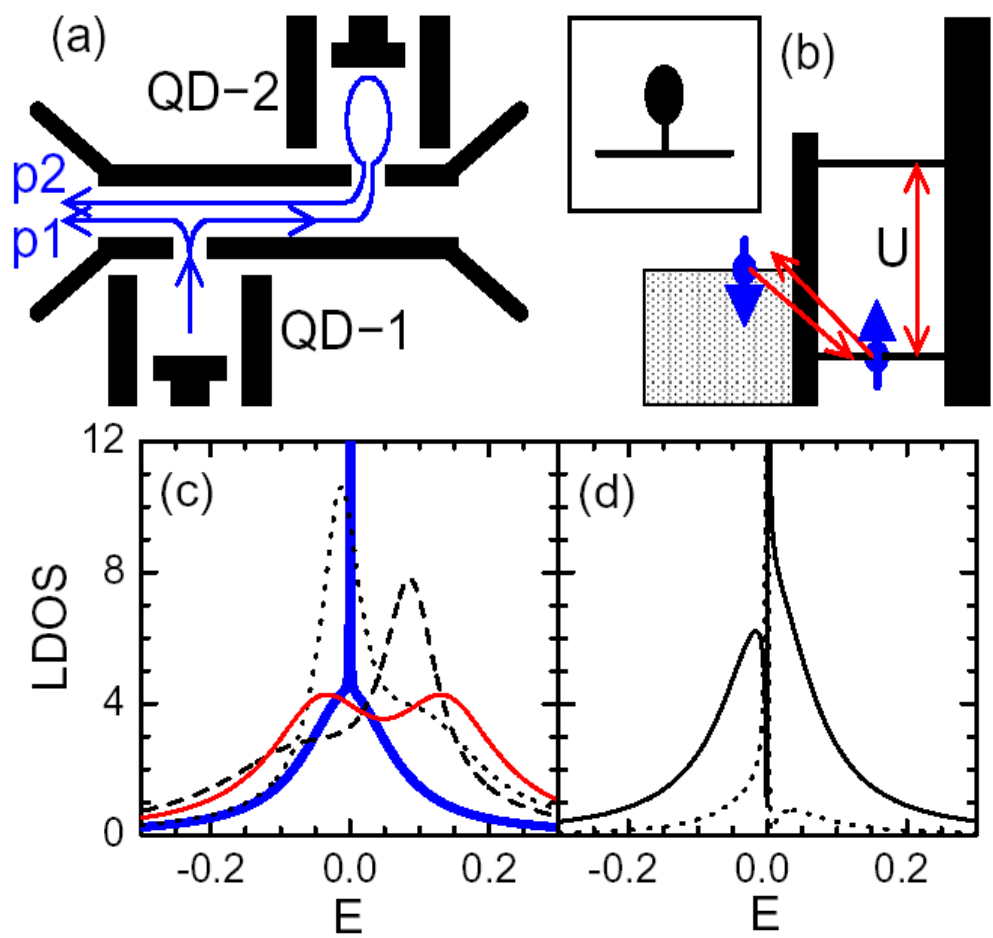
\begin{figure}
\caption{ (a). Schematic diagram for the device where two QDs
side-couple to a quantum wire. (b). Schematic diagram for a single
side-coupled QD and its co-tunneling processes.  (c). LDOS vs $E$
for $\epsilon_1=\epsilon_2=-2.3$ and $k_F L=\frac{\pi}{3}$ (dotted
curve), $\frac{\pi}{2}$ (thin solid curve), $\frac{2\pi}{3}$
(dashed curve), and $0.99\pi$ (thick solid curve). (d). Left dot
LDOS for $k_F L =0.99\pi$, $\epsilon_{1/2}=-2.3\pm
\Delta\epsilon/2$, and $\Delta \epsilon =-0.02$ (dotted curve) and
0.02 (solid curve). Temperature ${\cal{T}} =10^{-6}$ in (c) and
(d). } \label{fig1}
\end{figure}

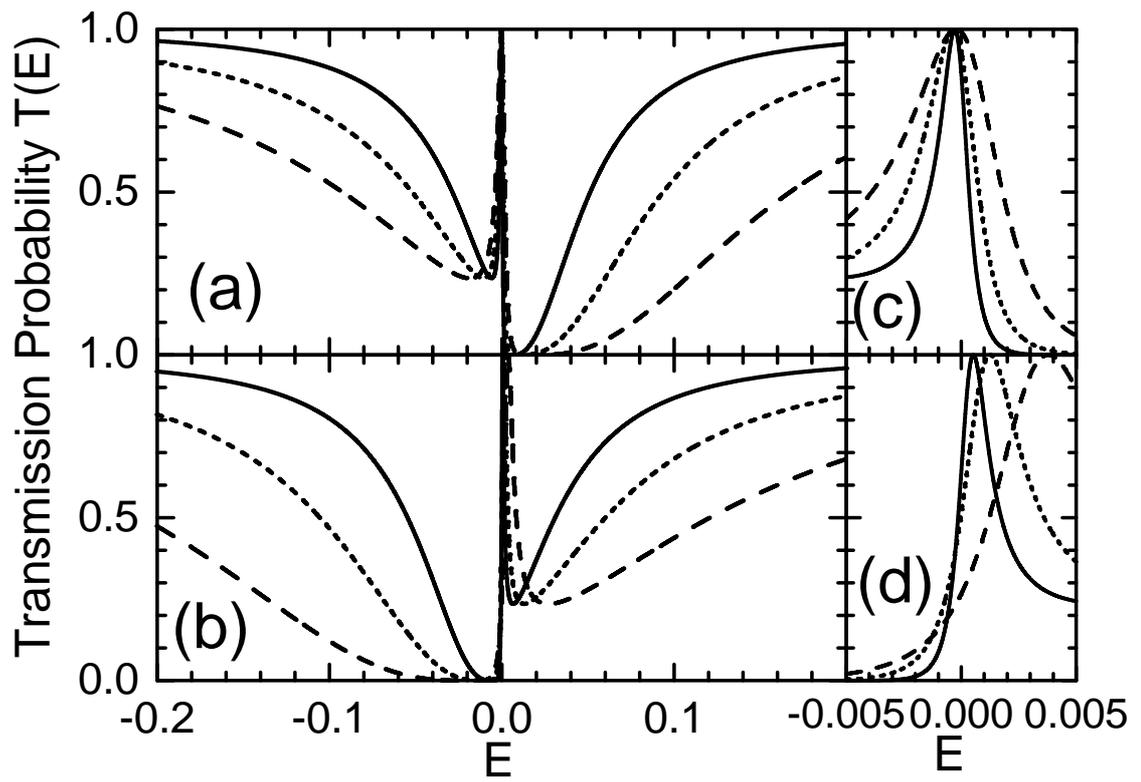
\begin{figure}
\caption{ $T(E)$ vs $E$ for ${\cal{T}}=10^{-6}$. The parameters
are: $\epsilon=-2.1$ (dashed curve), $-2.3$ (dotted curve), and
$-2.5$ (solid curve); $k_FL =\frac{\pi}{12}$ (a) and
$\frac{11\pi}{12}$ (b). (c) and (d) amplify the three curves of
(a) and (b) near $\mu =0$. }
\label{fig2}
\end{figure}

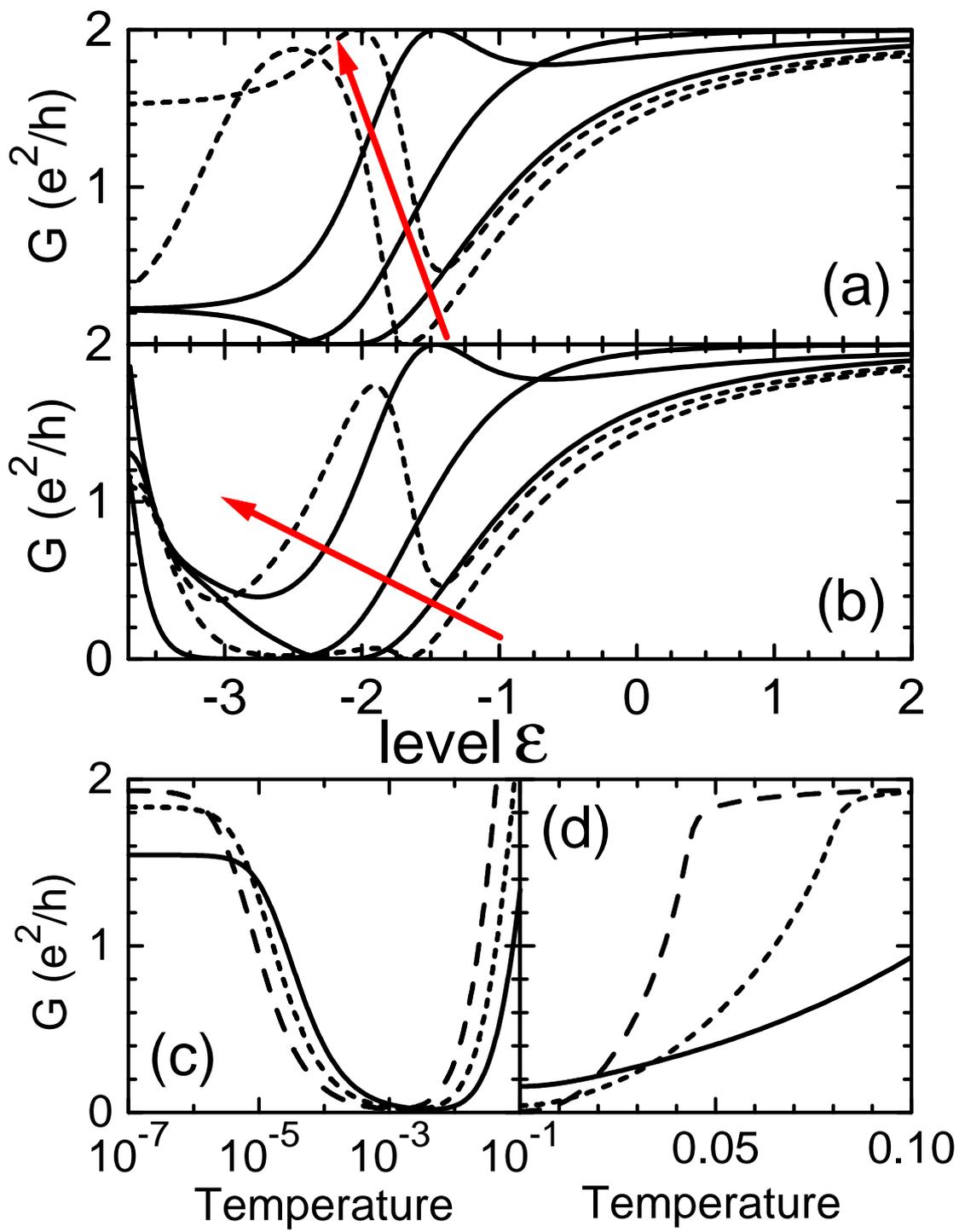
\begin{figure}
\caption{ (a) and (b): Conductance $G$ vs $\epsilon$ for
${\cal{T}}=10^{-6}$ (a) or $10^{-3}$ (b). $k_F L$ along the arrow
are for $0.99\pi$, $\frac{3\pi}{4}$, $\frac{\pi}{2}$,
$\frac{\pi}{4}$, and $\frac{\pi}{12}$, respectively. (c) and (d):
$G$ vs $\cal{T}$ for $k_F L =0.99\pi$ (c) or $\frac{\pi}{2}$ (d)
with $\epsilon=-2.1$ (solid curve), $-2.3$ (dotted curve), and
$-2.5$ (dashed curve), respectively. } \label{fig3}
\end{figure}

\end{document}